\documentclass[epj]{svjour}
\usepackage{cite,amsmath,eucal,epsfig}
\newcommand{\ud}{\mathrm{d}}
\begin{document}
\title{Microscopic systems with and without Coulomb interaction,
fragmentation and phase transitions in finite nuclei}
\author{J.M.~Carmona\inst{1}\thanks{\emph{carmona@df.unipi.it}}
\and J.~Richert\inst{2}\thanks{\emph{richert@lpt1.u-strasbg.fr}}
\and P.~Wagner\inst{3}\thanks{\emph{pierre.wagner@ires.in2p3.fr}}
}                     % Do not remove
\institute{Dipartimento di Fisica dell'Universit\`a and INFN,
I-56127 Pisa, Italy 
\and Laboratoire de Physique Th\'eorique,
Universit\'e Louis Pasteur,\\
3, rue de l'Universit\'e, 67084 Strasbourg Cedex (France) - UMR 7085
\and Institut de Recherches Subatomiques,
BP28, 67037 Strasbourg Cedex 2 (France) - UMR 7500}
%
%\date{\today}
\abstract{
  We test the influence of the Coulomb interaction on the thermodynamic
and cluster generation properties of a system of classical particles
described by different lattice models. Numerical simulations show that the
Coulomb interaction produces essentially a shift in temperature of quantities
like the specific heat but not qualitative changes. We also consider a
cellular model. The thermodynamic properties of the system are
qualitatively unaltered.
 % end of PACS codes
} %end of abstract
\authorrunning{Carmona, Richert, and Wagner}
\titlerunning{Microscopic systems with and without Coulomb interaction...}
\maketitle
\section{Introduction}
\label{intro}
\indent

  In the recent past the success of percolation models \cite{Richert} and 
their link with other generic approaches like Ising and Potts 
\cite{Schultz, Potts} has led to the development of many lattice models 
\cite{Samaddar, Samaddar1, DasGupta, Elattari, Samaddar2, 
Campi, Carmona, Borg, Chomaz} 
which were used as sensible albeit schematic descriptions of excited 
disassembling nuclei. As simple as they may appear, their thermodynamic 
properties were considered as being at least qualitatively those of bound 
nucleon systems which interact essentially by means of the short range nuclear 
interaction. It was tacitly implied that quantum effects do not qualitatively 
alter those properties, at the excitation energies which characterize 
fragmenting nuclei. 

  The introduction of time-independent stationary descriptions of 
fragmented systems presupposes that these systems are in thermodynamic 
equilibrium. Whether this is a realistic assumption and is effectively 
realised is still an open question. There exist however many data 
which show that most experimental results can be understood in this 
framework \cite{Richert}.

  The present contribution follows a double objective. First, in the 
framework of lattice models, we want to analyse the effect of the Coulomb 
interaction which is superimposed onto the short range nearest neighbour 
interaction which mimics the nuclear potential. This is done in the 
framework of the so called Ising Model with Fixed Magnetization (IMFM)
\cite{Carmona} and its extension to the case where proton-neutron are 
differentiated from proton-proton and neutron-neutron interactions. The 
treatment of the long range interaction is performed on the same footing as 
the short range one i.e. without any approximation in the calculations of 
the interaction between a given proton and all the others in the system.
In a second part we relax the lattice structure of the system.
Fragmenting nuclei are disordered systems whose constituents are 
not located at fixed positions on a regular lattice like in crystals but 
more in the continuum of position space. A priori, a more realistic 
description of such systems is realised in the framework of so called 
cellular models \cite{Elattari}. We show and discuss whether the freedom to 
occupy any space position leads or not to qualitative changes in the 
thermodynamic and topological (fragment formation) properties of the system.
  
  In section 2 we present a sketchy description of the IMFM model and the 
way we implement the Coulomb interaction which is close in form with recent 
work by Samaddar and Das Gupta \cite{Samaddar2}. We present and discuss the 
caloric curve, specific heat and phase diagram for undifferentiated and 
differentiated protons and neutrons. In this framework we work out the 
cluster content of the system and related observables for different values 
of the temperature. Most of the simulations are performed in the framework 
of the canonical ensemble. A comparison with microcanonical calculations 
is also presented. In section 3 we introduce the cellular model and study 
its thermodynamic properties which will be confronted with those obtained 
with the lattice model. Section 4 is devoted to comments and conclusions.

\section{Effect of the Coulomb interaction on finite systems described by 
Ising-type lattice models}
\label{sec:2}

\subsection{The models and their implementation}
\label{sec:21}
\indent

  The classical system with $A$ particles located on a cubic lattice with 
$L^3$ sites is described by Hamiltonians of the form
\begin{equation} H =-\sum_{<i,j>} V_{K_i K_j} n_i n_j + \frac{1}{2}
\sum_{i \neq j} \frac{e^2 Z_i Z_j}{r_{ij}} \label{Ham} \end{equation}
where $K_i \ (i=1,\cdots, L^3)$ labels either a proton (p) or a neutron (n). 
The short range potential $V_{K_i K_j}$ acts only between pairs of nearest 
neighbour particles $<i,j>$, when the sites $i$ and $j$ are both occupied
$(n_i = n_j = 1)$, unoccupied sites correspond to $n_k=0$. The interaction 
strengths are chosen in two different ways, either 
$V_{nn}=V_{pp}=V_{np}=\epsilon= 5$~MeV (undifferentiating potential, case
U in the sequel) or $V_{nn}=V_{pp}=0, \, V_{np}=\epsilon_{np}= 5.33$~MeV 
(differentiating potential, case D in the sequel). The latter, a priori more 
realistic choice has been used by different authors \cite{Borg, Chomaz, Pan,
Chomaz1}, the choice of strength is such that the classical ground state 
energy of finite nuclear systems is reproduced \cite{Lenk}.

  The Coulomb term in (\ref{Ham}) is such that $Z_i=0$ or 1 if particle $i$ 
is a neutron or a proton respectively. The distance between particles $i$ 
and $j$, $r_{ij}$, is determined on a lattice in which the distance between 
sites is fixed to $d=1.8$~fm.

  The canonical partition function reads
\begin{equation} \mathcal{Z}(\beta) = \sum_{[n_i,Z_i]} e^{-\beta H} \cdot
\delta_{ \sum_i n_i,A} \cdot \delta_{ \sum_i Z_i,Z} \end{equation} 
where $Z$ is the total number of charges in the system. In the 
calculations $Z/A$ was fixed at the value of 0.4.

  The thermodynamic properties and the space occupation by particles and 
bound clusters are obtained from realisations generated by means of 
Metropolis Monte Carlo simulations in which particles are moved on the 
lattice. Technical details about the algorithm used in the framework of the 
canonical ensemble have been given elsewhere \cite{Carmona}. We selected
$5 \times 10^4$ realisations corresponding each to $10 \times L^3$ 
Metropolis steps.

  We have also performed calculations in the framework of the 
microcanonical ensemble. A given realisation $(r)$ with fixed energy $E$ 
has a weight
\begin{equation} W_E(r) \sim (E-U(r))^{\frac{DA}{2} - 1} \cdot
\Theta(E-U(r)) \end{equation} 
where $U(r)$ is the potential energy, $\Theta$ is the step function, $D$ 
the space dimension, here $D=3$. Detailed balance fixes the 
acceptance rate from a realisation $(r)$ to a realisation $(r')$ to
\begin{equation} W_{r \to r'} = \min \bigg[ 1,\frac{W_E(r')}{W_E(r)} \bigg]\ . 
\label{bal} \end{equation}

  The simulations were done with open boundary conditions on the edges of 
the lattice both in the case of the canonical and microcanonical ensembles.

\subsection{Thermodynamics and cluster size distributions}
\label{sec:22}
\indent

  The quantities which we consider here are the caloric curve and the 
specific heat for fixed volume which in the canonical ensemble reads
\begin{equation} C_V = \frac{\ud <U>}{\ud T} = 
\frac{1}{T^2} \big( <U^2> - <U>^2 \big) \ . \nonumber \end{equation}
The brackets stand for an average over an ensemble of systems. In the 
microcanonical ensemble the temperature is related to the kinetic energy 
$K$ of the system by
\begin{equation} T = 2<K>/AD \ . \nonumber \end{equation}  
Then the specific heat can be cast in the simple form \cite{Ray}
\begin{equation} C_V = \frac{DA}{2} \bigg[ \frac{DA}{2} -
\Big( \frac{DA}{2} - 1 \Big) <K> \Big<\frac{1}{K} \Big> \bigg]^{-1} -DA \ .
\nonumber \end{equation}

  In the absence of the Coulomb interaction clusters of bound particles  
are determined in the same way as in ref. \cite{Campi} by application of 
the Coniglio-Klein prescription \cite{Coniglio} which fixes the condition 
under which a particle located on a site which is topologically connected 
to a cluster does effectively belong to this bound cluster. The test 
between nearest neighbour pairs of particles is made by means of the 
probability $p=1-\exp(-\epsilon/2T)$ in case U, with $\epsilon$ replaced by 
$\epsilon_{pp}, \epsilon_{nn}, \epsilon_{np}$ in case D, depending on the 
nature of the particles which are involved. The probability $p$ is then 
compared to a random number $\xi \in [0,1]$. If $\xi \le p$ the particles 
are considered as being bound, if $\xi > p$ they are not.

  The Coulomb interaction is taken into account in the case U and for a 
pair of neighbouring charged particles by replacing $\epsilon$ by 
$\epsilon - e^2/d$ where $d$ is the (fixed) distance between the particles. 
In the case D $p$ is not modified since neighbouring charged particles are 
not bound $\epsilon_{pp} = 0$. This procedure deserves some comments. The
identification of fragments by means of the Coniglio-Klein prescription is
conceptually valid if the interaction is short ranged, i.e. concerns only
nearest neighbours. This is not the case with the Coulomb interaction which 
acts over a large range. We restricted the tests with Coulomb for the
connection between particles to nearest neighbours. 
There is no real justification for this, since it is not possible 
to invoke screening effects, the classical Debye-H\"uckel screening length 
being much larger than the size of the system. Hence, as it stands, we have
no real means to estimate the validity of our procedure, except for 
quantitative arguments, i.e. the effect is certainly weak since 
$e^2/d \simeq 0.8$ MeV should be compared to $\epsilon \simeq 5$ MeV and
further the Coulomb interaction acts only on protons in a system where 
$N/Z = 1.5$ ($N =$ number of neutrons).

  The cluster size distribution can be determined for any fixed density 
$\rho = A/L^3$ and temperature $T$. From its knowledge one can work out 
different observables. Here we restrict ourselves to the behaviour of the 
largest cluster $A_{max}$ as a function of $\log S_2$, with $S_2=m_2/m_1$ 
and $m_k$ the $k^{th}$ moment of the distribution \cite{Campi1, Campi2}.

\subsection{Effect of the Coulomb interaction on the thermodynamic 
properties of the system}
\label{sec:23}
\indent

\begin{figure}[htb!]
\begin{center}
\epsfig{figure=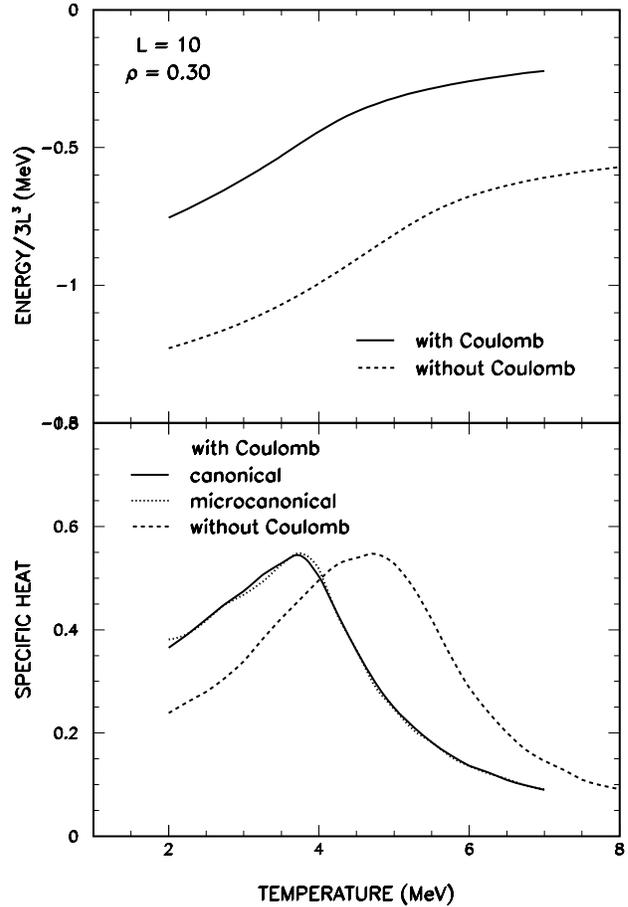,width=100mm}
\caption{Caloric curves and specific heat for a lattice model with
$A=300$ undifferentiated particles (protons and neutrons) in a volume
$L^3=10^3$. Full and open lines correspond respectively to results
including or not the Coulomb interaction. The specific heat calculated with
the Coulomb interaction in the framework of the microcanonical ensemble is
represented by a dotted line.}
\label{fig:1}
\end{center}
\end{figure}

\begin{figure}[htb!]
\begin{center}
\epsfig{figure=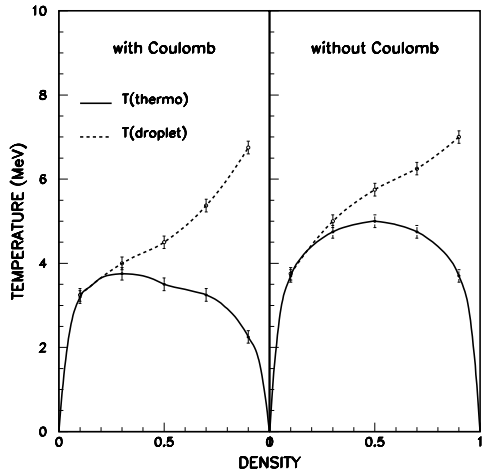,width=80mm}
\caption{ Phase diagrams $(\rho,T)$ with Coulomb interaction (left) and 
without Coulomb interaction (right) for undifferentiated particles. The full 
line indicates the phase separation, the dashed line the separation line 
between systems with heavy clusters (below) and light clusters and particles 
(above). See text.}
\label{fig:2}
\end{center}
\end{figure}

\begin{figure}[htb!]
\begin{center}
\epsfig{figure=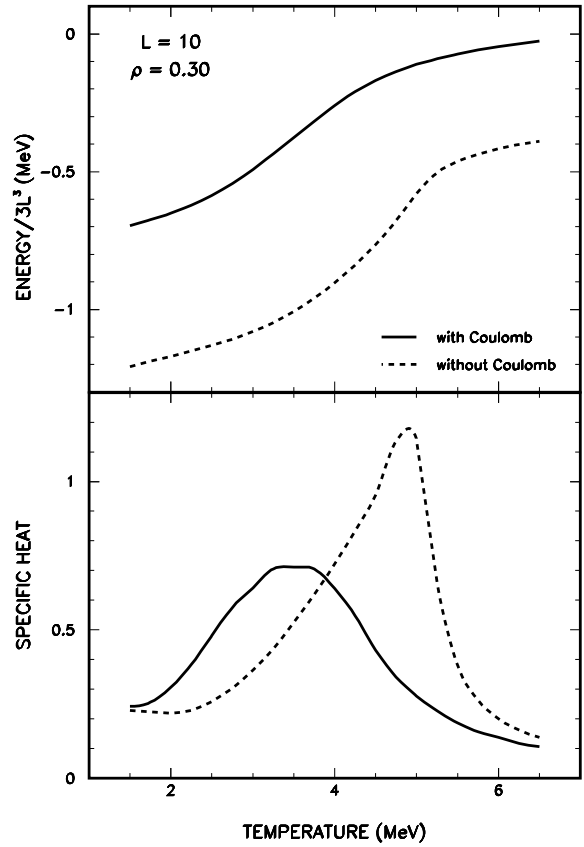,width=140mm}
\caption{Same as Fig.~\ref{fig:1} for differentiated particles (protons and
neutrons). Calculations are done in the canonical ensemble.}
\label{fig:3}
\end{center}
\end{figure}

  The effect of the Coulomb interaction has been studied in both cases U and
D with the two-body strengths given in 2.1. Fig.~\ref{fig:1} shows the caloric 
curve and the specific heat associated to a system with $L = 10$ and 
$\rho = 0.3$ for case U.
Calculations have been performed in both the canonical and 
microcanonical framework. As it can be seen in Fig.~\ref{fig:1}, on the 
specific heat, the results are undistinguishable. The point of interest is the
fact that the Coulomb interaction induces a sizable shift in the energy for 
fixed temperature on the caloric curve which increases approximately 
parallel to each other and a reduction of about 1.5~MeV in the 
temperature corresponding to the maximum of $C_V$. The observed reduction
of the temperature can be understood by means of the following arguments.
The Coulomb interaction takes a part of the total available
energy and hence the thermal energy is decreased leading to a decrease of 
the temperature. With changing excitation energy the Coulomb energy does not
change much, hence the energy differences between the cases without and 
with Coulomb interaction stay approximately constant. This behaviour is 
general, whatever the size of the system and its density. The expected drop 
in the temperature in the presence of Coulomb reflects also in the $(\rho,T)$
phase diagram shown in Fig.~\ref{fig:2}, in which the coexistence line has 
been fixed by following the maximum of $C_V$. One may mention the slight 
dissymmetry with respect to $\rho=0.5$. For larger $\rho$ the Coulomb 
effect is larger than for small densities and consequently produces a 
stronger shift in the temperature which defines the border of the 
coexistence region. This is due to the fact that higher $\rho$ 
corresponds to stronger packing, hence charges interact more effectively.
In case D, when the particles are differentiated, the results are 
qualitatively similar for both the caloric curve and the specific heat with 
a similar shift of 1.5~MeV in the temperature, see Fig.~\ref{fig:3}.
In this case, the maximum of $C_V$ is much more damped when Coulomb is 
present than in case U, and the transition region is therefore less 
precisely defined.

  One may notice that in both cases, with and without Coulomb interaction,
the caloric curve is unable to give an indication about the order of the
transition. In former studies \cite{Carmona} it has been shown through the
attempt to determine critical exponents that the transition could be 
continuous, but it came out that the result was not conclusive because of 
the small size of the system used in order to apply finite size scaling 
arguments. Microcanonical calculations on small systems  \cite{Carmona1}
also seemed to indicate the continuous nature of the transition, no
backbending in the caloric curve could be observed. This point was later
discussed in ref.\cite{Chomaz} and recently it was shown by Pleimling and
H\"uller \cite{Pleimling} that the backbending can be observed if the 
system gets large enough, which shows that the transition is in fact first
order when one goes to the thermodynamic limit.

\subsection{The effect of the Coulomb interaction on the cluster size 
distributions and related observables}
\label{sec:24}
\indent

\begin{figure}[htb!]
\begin{center}
\epsfig{figure=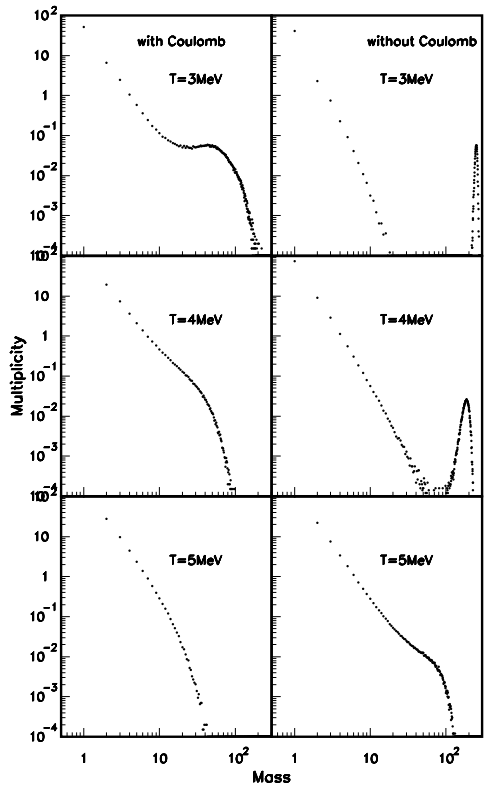,width=120mm}
\caption{Cluster size multiplicities for different temperatures. Left
column: with Coulomb interaction. Right column: without Coulomb
interaction. See text.}         
\label{fig:4}
\end{center}
\end{figure}

  The behaviour of the mass distribution of clusters is shown in 
Fig.~\ref{fig:4} for case D, 
temperatures lie in the range $3 - 5$~MeV. One observes again 
a shift in the behaviour of the system when the Coulomb interaction is 
switched on. For the same temperature, the distribution without Coulomb 
contains more larger clusters. This is understandable since the long range 
interaction is repulsive and tends to split the bond fragments. The whole 
picture is also consistent with the thermodynamic behaviour of the system. 
This is seen in Fig.~\ref{fig:2} on the line which separates the heavy cluster
behaviour below the line from the light cluster behaviour above it and 
stands as the finite size remnant of a continuous set of transition points 
in the percolation framework \cite{Stauffer}. Indeed, the line lies at 
lower temperatures when the Coulomb interaction is present. The 
Coniglio-Klein procedure is the correct prescription in order to identify 
bound clusters of interacting particles with kinetic energy. It ensures 
that a given particle is bound to a cluster or not by means of separation 
energy arguments \cite{Campi}. One would expect that this separation line 
ends at $\rho = 0.5$ on the coexistence line. This is not the case here. 
It may be due to the criterion (the maximum of $C_V$) which is used here
in order to fix this coexistence line. It clearly corresponds to a finite 
size effect. Indeed we checked that the distance between the separation 
line and the $\rho = 0.5$ point on the coexistence line gets smaller and
smaller when the size of the system increases.

\begin{figure}[htb!]
\begin{center}
\epsfig{figure=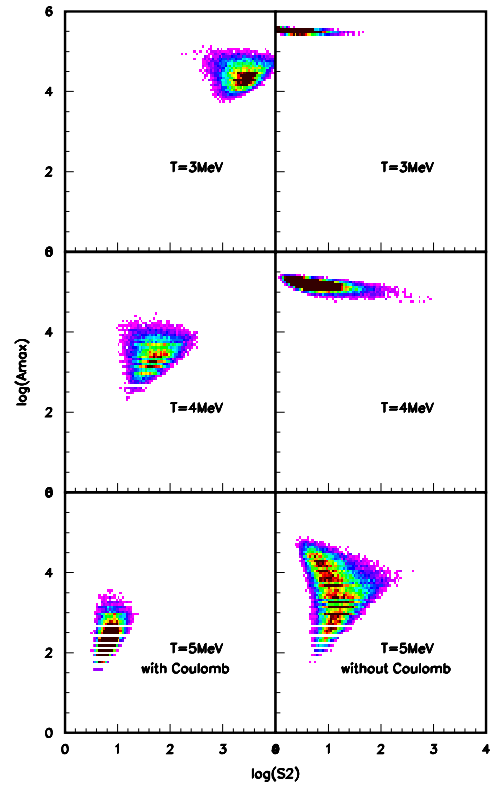,width=120mm}
\caption{ Size of the heaviest cluster $A_{max}$ vs. $S_2=m_2/m_1$ 
represented for the different Monte Carlo events. $m_2$ and $m_1$ are the 
second and first moments of the cluster size distribution.}
\label{fig:5}
\end{center}
\end{figure}

Fig. \ref{fig:5} shows the correlation between $A_{max}$ and $S_2$ in case D. 
Events which come at large $A_{max}$ correspond to the presence of heavy 
clusters, those with small $A_{max}$ to the presence of many small clusters. 
The intermediate region shows events which correspond to the transition 
region in the percolation framework. As it can be seen, this region appears 
again for lower temperatures in the case where the Coulomb interaction is 
acting.

\begin{figure}[htb!]
\begin{center}
\epsfig{figure=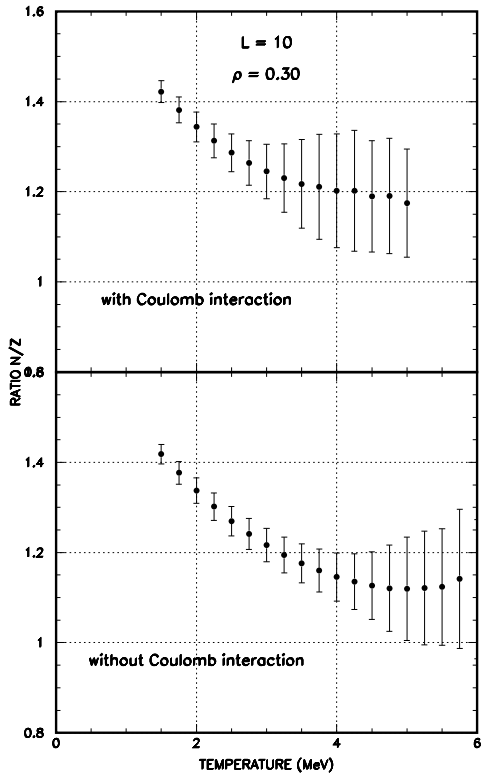,width=120mm}
\caption{Average isotopic ratios $N/Z$ and fluctuations widths in the 
clusters as a function of the temperature in the presence and absence of 
the Coulomb interaction. Volume $L^3=10^3$, density $\rho = 0.3$. 
See comments in the text.}
\label{fig:6}
\end{center}
\end{figure}

  A further observable of physical interest which can be experimentally 
determined is the ratio of the number of neutrons to the number of protons, 
$N/Z$, which are present in the clusters \cite{Pochodzalla}. 
Fig.~\ref{fig:6} shows the 
evolution of this ratio for clusters of mass $A=N+Z \geq 40$ with increasing 
temperature for case D. One observes a decrease of the relative number of 
neutrons, both in the absence and the presence of the Coulomb interaction. 
This is in agreement with the results of ref. \cite{Chomaz1} and, for heavy
clusters, with the experimental results of ref. \cite{Milazzo}. The
decrease is apparently slightly more pronounced in the case where the 
Coulomb interaction is active, although, as indicated, the fluctuations from 
event to event increase with increasing temperature. This result may appear 
to be somewhat paradoxical. It may be due to the fact that for fixed 
temperature the cluster size distribution is different in both cases as we 
already saw above.

\begin{figure}[htb!]
\begin{center}
\epsfig{figure=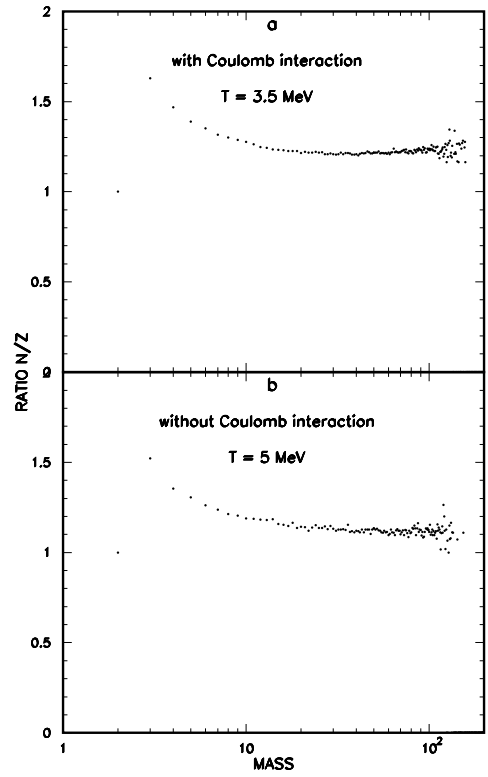,width=120mm}
\caption{ Isotopic ratios $N/Z$ in the clusters as a function of the
cluster size. Upper part (a): with Coulomb interaction. Lower part (b): 
without Coulomb interaction.}
\label{fig:7}
\end{center}
\end{figure}

  Finally, the evolution of the $N/Z$ ratios with the mass is shown in 
Figs.~\ref{fig:7} for case D. In Fig.~\ref{fig:7}a the results correspond 
to the case where the Coulomb interaction is included. The ratio drops 
from $N/Z \sim 1.6$ for $A=3$ to 1.2 for the largest masses. A similar 
behaviour is observed in Fig.~\ref{fig:7}b for a higher temperature and in 
the absence of the Coulomb interaction. 
Hence lighter species are more neutron rich than heavier ones. 
This fact which has already been observed in the calculations of ref. 
\cite{Chomaz1} in the absence of the Coulomb interaction remains valid when 
Coulomb is present and seems to be in agreement with the experimental 
findings \cite{Yennello, Xu}.

\section{Cellular model approach to the description of fragmenting nuclei}
\label{sec:3}
\indent

  A system of $A$ particles is contained in a cube of volume $V$ in $3D$
space and divided into cells of volume $d^3, \ V = L^3 d^3$. Cells are 
either empty or occupied by at most one particle characterized by its random 
position $\vec r_i \ (i=1, \cdots ,A)$ in the cell and random momentum
$\vec p_i$ \cite{Elattari}. The particles are classical. Neighbouring 
particles interact through a short range two-body potential $V_0(r_{ij})$ 
which is repulsive at short distance and attractive for 
$r_{ij} \ge 1.55$~fm \cite{Wilets}. 
In the calculations $d = 1.8$~fm. The Coulomb 
interaction is not taken into account. The Hamiltonian is written as
\begin{equation} H = \sum_{i=1}^A \frac{p_i^2}{2m} 
- \sum_{<ij>}V_0(r_{ij}) n_i n_j \end{equation}
with $n_i=0(1)$ if the cell $i$ is empty (occupied).

  Similarly to the lattice case we have worked out the thermodynamic 
properties of such a system in the framework of both the canonical and 
microcanonical ensemble. The generation of configurations by means of a 
Metropolis Monte Carlo procedure is performed in the following way. 
Starting from an initial configuration in which the particles are disposed 
randomly into the available cells one performs either the move of a 
particle into an empty cell, or changes the position of a particle in its 
cell. Each operation is performed with a probability $1/2$, the application 
of the Metropolis Monte Carlo algorithm leads to a minimum in energy in the 
framework of the canonical ensemble. In the case of the microcanonical 
ensemble the test is the one described in section \ref{sec:21}, 
Eq. (\ref{bal}).

\begin{figure}[htb!]
\begin{center}
\epsfig{figure=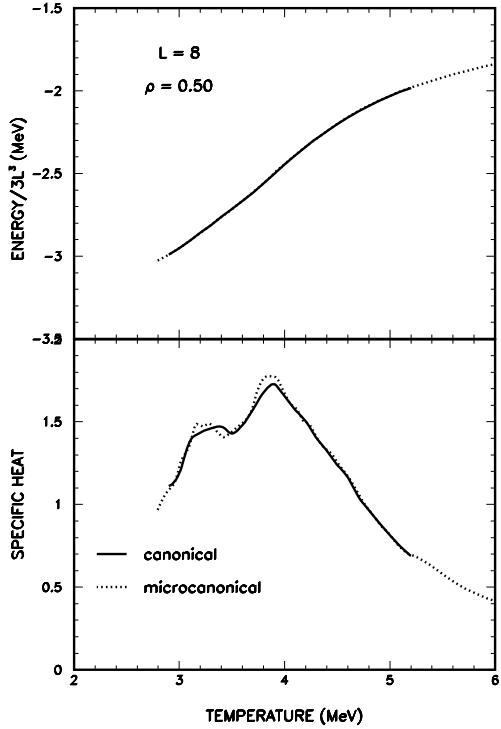,width=120mm}
\caption{Canonical and microcanonical caloric curves and specific heat in
the framework of the cellular model for a system of $A=256$ particles in a
volume $L^3=8^3$. See comments in the text.}
\label{fig:8}
\end{center}
\end{figure}

\begin{figure}[htb!]
\begin{center}
\epsfig{figure=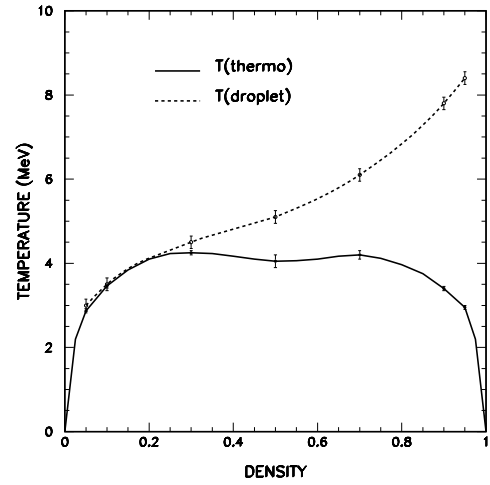,width=80mm}
\caption{Phase diagrams $(\rho,T)$ in the framework of the cellular model
for a system with $A=256$ particles in a volume $L^3=8^3$. The dashed line
has the same meaning as in Fig.~\ref{fig:2}.}
\label{fig:9}
\end{center}
\end{figure}

  Typical results can be seen on Fig.~\ref{fig:8} which shows the caloric curve
and the specific heat. Both ensembles lead to a very close behaviour of 
these quantities. This behaviour is qualitatively similar to those obtained 
with lattice models. The same is true for the corresponding phase diagram 
shown in Fig.~\ref{fig:9}. The thermodynamic phase coexistence line shows two
slight maxima symmetrical with respect to the critical point at the density
$\rho = 0.5$. This behaviour may be due to the way in which we define the
coexistence line (maximum of the specific heat) and (or) related to finite 
size effects, see ref.\cite{Carmona}.
The present calculations have been performed for a fixed 
volume with $8^3$ sites and the possible transition looks continuous in 
this case. The small depression for $\rho=0.5$ reminds the same effect which 
was found in the IMFM calculations of ref. \cite{Carmona}. The dotted line 
shows the separation line of heavy and light clusters as in the case of the 
lattice models. Its behaviour is very similar to the one observed in 
Fig.~\ref{fig:2}. The fact that its lower end does not coincide with 
$\rho=0.5$  may have the same reasons than those presented in 
section \ref{sec:24}.

\section{Summary and conclusions}
\indent

  In the present work we aimed to present and discuss two points related to 
the description of nuclear fragmentation by means of microscopic classical 
models.

  We first looked for the effects induced by the long range Coulomb 
interaction which acts between charged particles in the presence of a short 
range attractive potential mimicking the nuclear interaction. We have 
shown on a lattice model that the Coulomb interaction does not induce any 
qualitative change in the thermodynamics and cluster size distribution in 
the different phases in which the system exists. The essential quantitative 
effect is a global systematic and sizable shift of the temperature of the 
system by about 1.5~MeV in the models which have been worked out both in 
the thermodynamic and cluster transitions. The same type of effect has 
already been mentioned in former studies \cite{Biro}. We have also 
investigated the behaviour of the cluster content, both in the absence and 
presence of the Coulomb interaction.

  In a second part we introduced a cellular model aimed to describe a 
disordered system of particles in thermodynamic equilibrium in the 
framework of the canonical and microcanonical ensembles. The results show 
that such models, though a priori more realistic than lattice models, are 
descriptions which lead qualitatively to the same properties, at least when 
finite systems are concerned. This leads to the conclusion that lattice 
models which are simple to handle should be good enough for a schematic 
description of nuclear fragmentation processes if the generated systems are 
in thermodynamic equilibrium. There remains however the problem concerning 
the importance of quantum effects. These may be the weaker the higher the 
temperature, but this point has to be confirmed.

\end{document}